\documentclass[apjl]{emulateapj}
\slugcomment{Received 2013 October 28; 
accepted 2013 December 10; published 2014 January 6}
\usepackage{psfig}
\usepackage{natbib}
\usepackage{hyperref}
\hypersetup{
    colorlinks=true,
    linkcolor=blue,
    citecolor=blue}

\newcommand{\ssst}{\scriptscriptstyle}
\newcommand{\E}[1]{\times 10^{#1}}
\newcommand{\etal}{et al.}

\newcommand{\decl}[3]{{#1}^{\circ}{#2}'{#3}''}
\newcommand{\RAdot}[3]{{#1}^{{\rm h}}{#2}^{{\rm m}}{#3}\fs}
\newcommand{\decldot}[3]{{#1}^{\circ}{#2}'{#3}\farcs}

\newcommand{\s}{\,{\rm s}}      \newcommand{\ps}{\,{\rm s}^{-1}}
    
\newcommand{\cm}{\,{\rm cm}}    \newcommand{\km}{\,{\rm km}}
\newcommand{\kms}{$\km\ps$}
\newcommand{\kpc}{\,{\rm kpc}} 
\newcommand{\erg}{\,{\rm erg}}

        \newcommand{\NH}{N_{\ssst\rm H}}
\newcommand{\no}{n_{\ssst 0}}

\newcommand{\ROSAT}{{\sl ROSAT}} 

\newcommand{\XMMN}{{\sl XMM-Newton}}
\newcommand{\Swift}{{\sl Swift}}
\newcommand{\Chandra}{{\sl Chandra}}

\newcommand{\snr}{Kes~79}
\newcommand{\src}{3XMM~J185246.6+003317}

\shorttitle{}

\begin{document}

\title{Discovery of the transient magnetar \src\ near 
supernova~remnant~Kesteven~79 with \XMMN}

\author{
Ping Zhou\altaffilmark{1,2},
 Yang Chen\altaffilmark{1,3},
 Xiang-Dong Li\altaffilmark{1,3},
 Samar Safi-Harb\altaffilmark{2,4},
 Mariano Mendez\altaffilmark{5},
 Yukikatsu Terada\altaffilmark{6},
 Wei~Sun\altaffilmark{1},
 and Ming-Yu Ge\altaffilmark{7}
}

\altaffiltext{1}{Department of Astronomy, Nanjing University, 
Nanjing~210093, China}
\altaffiltext{2}{Department of Physics and Astronomy, University of 
Manitoba, Winnipeg, MB R3T 2N2, Canada}
\altaffiltext{3}{Key Laboratory of Modern Astronomy and Astrophysics,
 Nanjing University, Ministry of Education, China}
\altaffiltext{4}{Canada Research Chair}
\altaffiltext{5}{Kapteyn Astronomical Institute, University of Groningen,
P.O. Box 800, 9700 AV Groningen, the Netherlands}
\altaffiltext{6}{Graduate School of Science and Engineering, Saitama University,
255 Simo-Ohkubo, Sakura-ku, Saitama 338-8570, Japan}
\altaffiltext{7}{Key Laboratory for Particle Astrophysics, Institute of 
High Energy Physics, Chinese Academy of Sciences, Beijing 100049, China}

\begin{abstract}
We report the serendipitous discovery with \XMMN\ that \src\ is an 11.56~s 
X-ray pulsar located $1'$ away from the southern boundary of supernova 
remnant \snr.
The spin-down rate of \src\ is $<1.1\E{-13}~\s\ps$, 
which, together with the long period $P=11.5587126(4)\s$, indicates a dipolar 
surface magnetic field of $<3.6\E{13}$~G, a characteristic age of 
$>1.7$~Myr, and a spin-down luminosity of $<2.8\E{30}\erg\ps$.
Its X-ray spectrum is best-fitted with a resonant Compton 
scattering model and can be also adequately described by a blackbody model.
The observations covering a seven month span from 2008 to 2009 show variations
in the spectral properties of the source,
with the luminosity decreasing from $2.7\E{34}~\erg\ps$ to $4.6
\E{33}~\erg\ps$, along with a decrease 
of the blackbody temperature from $kT\approx 0.8$~keV to $\approx0.6$~keV.
The X-ray luminosity of the source is higher than its spin-down 
luminosity, ruling out rotation as a power source.
The combined timing and spectral properties, the non-detection
of any optical or infrared counterpart, together with the lack of
detection of the source in archival X-ray data prior to the 2008 \XMMN\ 
observation, point to \src\ being a newly discovered transient low-B magnetar 
undergoing an outburst decay during the \XMMN\ observations.
The non-detection by \Chandra\ in 2001 sets an upper limit of $4\E{32}\erg\ps$
to the quiescent luminosity of \src.
Its period is the longest among currently known transient magnetars.
The foreground absorption toward \src\ is similar to that of \snr, 
suggesting a similar distance of $\sim 7.1$~kpc.

\end{abstract}


\section{Introduction}

\begin{figure*}
\centerline{ {\hfil\hfil
\psfig{figure=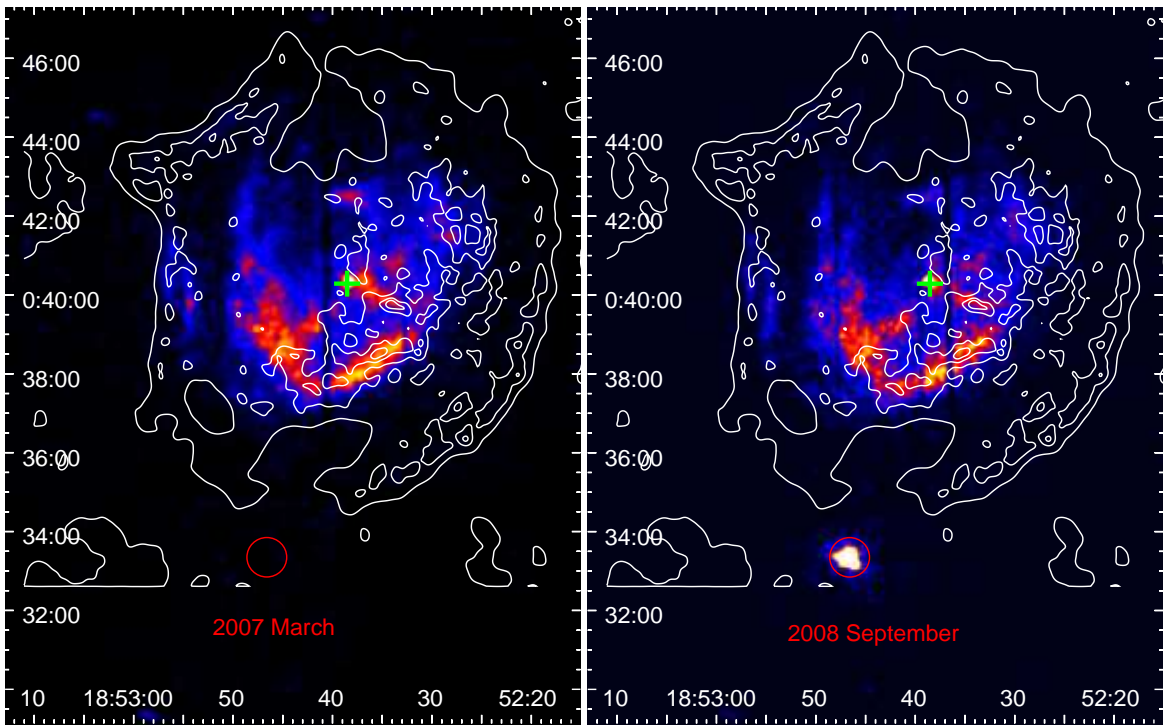,width=6.4in ,angle=0, clip=}
\hfil\hfil}}
\caption{
Raw EPIC-MOS2 image of \src\ and the northern SNR \snr\ from the
observations taken during 2007 March 20--21 (Obs ID 0400390301, left panel) and
2008 September 23 (Obs ID 0550670401, right panel), respectively.
The images have been smoothed using a Gaussian with $\sigma$=3 and
are shown with the same intensity scale.
Radio contours are overlaid using Very Large Array 1.4 GHz continuum emission.
The red circle and the green cross indicate the locations of
the transient magnetar and the CCO, respectively.
}
\label{f:transient}
\end{figure*}

Anomalous X-ray pulsars (AXPs) and soft gamma-ray repeaters (SGRs) have been
recognized as manifestations of a small class of known neutron stars 
dubbed ``magnetars'', commonly believed to be powered by the decay of their 
strong magnetic fields \citep{1995MNRAS.275..255T, 1996ApJ...473..322T,
2002ApJ...574..332T}.
Nearly two decades of observations show that this population of objects 
rotate slowly in comparison to the classical rotation-powered 
pulsars, with periods $P\sim 2$--12 s, large period derivatives $\dot{P} 
\sim 10^{-13}$--$10^{-10}~\s\ps$, and highly variable X-ray emission. 
Their X-ray luminosities are notably larger than their rotational energy
loss, implying that their powering mechanism cannot be rotation but 
rather magnetic field decay, with surface dipole magnetic fields 
(inferred from $P$ and $\dot{P}$) exceeding the quantum 
critical value $B_{\rm QED}= 4.4\E{13}$~G 
\citep[for reviews, see][]{2008A&ARv..15..225M, 2013BrJPh.tmp...38M,
2011arXiv:1101.4472, 2013IAUS..291...11R}.
Other models have been also proposed to explain magnetars, including
accretion from a fallback disk 
\citep{2013IAUS..290...93A} 
or quark stars 
\citep{2007A&A...473..357O}.
To date, there are only 26 known magnetars\footnote{http://www.physics.mcgill.ca/$\sim$pulsar/magnetar/main.html} 
\citep{2013arXiv1309.4167O}.
Despite being a small sample among the known neutron star population, magnetars 
have attracted wide and growing interest in the last
decade, among both the observational and theoretical communities, 
continually providing us with surprises and unexpected discoveries
that are shaping our understanding of the diversity of neutron stars and 
blurring the distinction between them.

This growing diversity includes, in addition to the magnetars and 
``classical'' rotation-powered pulsars, high-B radio pulsars, rotating 
radio transients, 
X-ray dim isolated neutron stars, 
and the central compact objects (CCOs), with the latter recently dubbed 
as ``anti-magnetars'' and showing evidence for a much lower magnetic field 
\citep[$\sim$10$^{10}$--10$^{11}$~G;][]{2013ApJ...765...58G}.
A number of recent observations showed that a super-critical 
dipole magnetic field ($B$$\geq$$B_{\rm QED}$) is not necessary for 
neutron stars to display magnetar-like behavior 
\citep[e.g., the discoveries of two low-B magnetars: SGR~0418$+$5729 and
Swift~J1822.3$-$1606;][]{2010Sci...330..944R, 2012ApJ...754...27R,
2012ApJ...761...66S},
that magnetars can be also radio emitters 
\citep[e.g.,][]{2006Natur.442..892C},
and that one high-B pulsar (PSR J1846$-$0258 in supernova remnant (SNR) 
Kes~75), thought to be an exclusively rotation-powered radio pulsar, 
behaved like a magnetar 
\citep{2008ApJ...678L..43K, 2008Sci...319.1802G}.
Moreover, the discovery of a handful of transient magnetars
\citep[i.e., previously missing magnetars discovered following an 
outburst typified by XTE~J1810--197;][]{2004ApJ...609L..21I}
are showing us that there is likely a large 
population of magnetars awaiting discovery.

Among the current known magnetars, eight have been so far identified as 
transient magnetars (XTE~J1810$-$197, AX~J1845$-$0258, 
CXOU~J164710.2$-$455216,  1E~1547$-$5408, SGR~1627$-$41, 
SGR~1745$-$29, SGR 0501+4516, and PSR~J1622$-$4950),
the study of which is providing a wealth of information about the emission 
mechanisms and evolution of these objects, as well as their connection 
to the other classes of neutron stars 
\citep{2011arXiv:1101.4472}.
Any new addition would be important to increase this small sample and 
provide physical insights on their physical properties.

In this Letter, we report on the \XMMN\ discovery of a transient low-B
magnetar located south of SNR Kes~79 hosting a CCO 
\citep{2003ApJ...584..414S}.
This source was serendipitously discovered during our multiwavelength 
study of Kes~79 
\citep[P. Zhou \etal\ in preparation;][]{2013arXiv1304.5367C},
adding a new member to the growing class of transient magnetars.

\section{Observations}
The observations described here were carried out with \textit{XMM-Newton} 
pointing at SNR~Kes~79 using the European Photon Imaging Camera (EPIC)
which is equipped with one pn 
\citep{2001A&A...365L..18S}
and two MOS cameras 
\citep{2001A&A...365L..27T}
covering the 0.2--12 keV energy range.

We found a bright point-like source just outside the southern boundary of 
the remnant in the 2008--2009 observations (see below) at around the 
following coordinates (J2000): 
R.A.\ =$\RAdot{18}{52}{46}6$ decl.\ =$\decldot{00}{33}{20}9$, located 
$7\farcm4$ away from the CCO (see Figure~\ref{f:transient}).
Checking against the \textit{XMM-Newton} Serendipitous Source 
Catalogue,\footnote{http://xmmssc-www.star.le.ac.uk/Catalogue/3XMM-DR4/ UserGuide$_{-}$xmmcat.html}
we find a point source (\src) which coincides with the 
above-mentioned source.
Therefore, we will refer to the source hereafter by its 3XMM name.

\begin{figure}
\centerline{ {\hfil\hfil
\psfig{figure=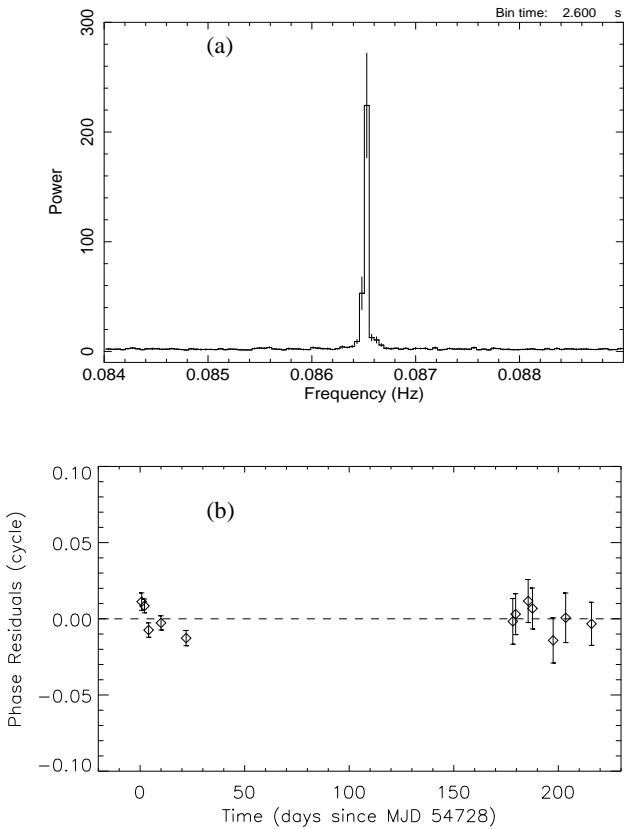,width=3.5in,angle=0, clip=}
\hfil\hfil}}
\caption{
(a): the averaged power spectrum for the 12 time series of \src\ 
reveals a highly significant period at around 0.0865~Hz (11.56~s). 
(b): phase residuals for the 12 observations after subtraction of 
best-fit model ($P=11.5587126 (4)\s$ at MJD 54728.75 and $\dot{P}=5.7\pm4.8
\E{-14}\s\ps$).
}
\label{f:fig2}
\end{figure}

\setlength{\tabcolsep}{1.2pt} 
\tabletypesize{\footnotesize}
\begin{deluxetable*}{lcccccccccccc}
\tablecaption{Summary of the 12 \XMMN\ epoch observations,
their timing and spectral properties}
\tablehead{
& & & & & & $power$-$law$ & & \multicolumn{2}{c}{$bbodyrad$} &
& \multicolumn{2}{c}{$RCS$} \\
\cline{7-7} \cline{9-10} \cline{12-13}
\colhead{ObsID}  & Obs. Date & Exposure$^a$ & Start Epoch$^b$ & Period$^c$
& $f_p$ &  $\Gamma$ & & $kT_{\rm bb}$ & $R_{\rm bb}$ & & $kT_{\rm rcs}$ & $F_X$ \\ 
& & (ks) & (MJD) & (s) & $\%$ & & & (keV) & (km) & & 
(keV) & ($10^{-12} \erg \cm^{-2} \s$)
}
\startdata
0550670201 \dotfill & 2008 Sep 19 & 21.6/21.6 & 54728.75  & 11.55856 (14)
& $62\pm7$
& $3.14\pm0.09$ &
& $0.80\pm0.02$ & $0.73\pm0.04$&
& $0.71^{+0.03}_{-0.06}$ & $4.7\pm1.0$ \\
0550670301 \dotfill & 2008 Sep 21 & 30.3/30.3 & 54730.07 & 11.55853 (7) 
& $64\pm6$
& $3.12\pm0.08$ &
& $0.80\pm0.02$ & $0.70\pm0.03$ &
& $0.71^{+0.02}_{-0.05}$ & $4.5\pm0.8$ \\ 
0550670401 \dotfill & 2008 Sep 23 & 35.4/35.4 & 54732.07 & 11.55879 (7) 
& $62\pm5$
& $3.16\pm0.08$ &
& $0.80\pm0.02$ & $0.73\pm0.03$ &
& $0.71^{+0.02}_{-0.06}$ & $4.6\pm 0.9$ \\
0550670501 \dotfill & 2008 Sep 29 & 33.3/33.3 & 54738.01 & 11.55865 (7) 
& $56\pm6$
& $3.12\pm0.08$ &
& $0.81\pm0.02$ & $0.74\pm0.03$ &
& $0.71^{+0.02}_{-0.04}$ & $5.0\pm0.8$ \\
0550670601 \dotfill & 2008 Oct 10 & 35.6/30.5 & 54750.01 & 11.55870 (7) 
& $68\pm6$
& $3.28\pm0.08$ &
& $0.78\pm0.02$ & $0.70\pm0.04$ &
& $0.68^{+0.03}_{-0.06}$ & $3.9\pm0.8$ \\
0550670901 \dotfill & 2009 Mar 17 & 26.2/23.3 & 54907.60 & 11.55854 (35) 
& $58\pm11$
& $3.83\pm0.24$ &
& $0.63\pm0.04$ & $0.50\pm0.07$ &
& $0.53\pm0.03$ & $0.8\pm0.2$ \\
0550671001 \dotfill & 2009 Mar 16 & 27.3/20.0 & 54906.25 & 11.55883 (39) 
& $71\pm10$
& $3.24\pm0.28$ &
& $0.73\pm0.06$ & $0.37\pm0.06$ &
& $0.62\pm0.07$ & $0.8\pm0.3$ \\
0550671101 \dotfill & 2009 Mar 25 &  18.9/0 & 54915.66 & 11.55876 (52) \\
0550671201 \dotfill & 2009 Mar 23 & 27.1/15.7 & 54913.58 & 11.55861 (35) 
& $75\pm10$
& $3.65\pm0.30$ &
& $0.67\pm0.06$ & $0.44\pm0.08$ & 
& $0.56\pm0.05$ & $0.8\pm0.3$ \\
0550671301 \dotfill & 2009 Apr 04 & 26.2/20.1 & 54925.54 & 11.55886 (36) 
& $76\pm10$
& $3.62\pm0.30$ &
& $0.63\pm0.05$ & $0.46\pm0.08$ &
& $0.54\pm0.05$ & $0.7\pm0.3$ \\
0550671801 \dotfill & 2009 Apr 22 & 28.2/28.2 & 54943.89 & 11.55903 (39) 
& $60\pm11$
& $3.81\pm0.28$ &
& $0.63\pm0.05$ & $0.46\pm0.07$ &
& $0.53\pm0.05$ & $0.7\pm0.2$ \\
0550671901 \dotfill & 2009 Apr 10 & 30.7/14.4 & 54931.53 & 11.55868 (37) 
& $77\pm12$
& $3.54\pm0.37$ &
& $0.65\pm0.07$ & $0.44\pm0.10$ & 
& $0.55\pm0.06$ & $0.7\pm0.3$
\tablecomments{$f_p$ is the pulsed fraction after background subtraction
(1$\sigma$ uncertainty).
$\Gamma$ is the photon index inferred from the $power$-$law$ model ($\chi_\nu
^2~({\rm d.o.f}) =1.46~(893)$);
$kT_{\rm bb}$ and $R_{\rm bb}$ are the temperature and radius obtained from 
the $bbodyrad$ model ($\chi_\nu^2~({\rm dof})=1.10~(893)$). 
$kT_{\rm rcs}$ is the temperature obtained from the $RCS$ model ($\chi_\nu^2~(
{\rm dof})=1.07~(891)$) for strongly magnetized sources. 
$F_X$ is the unabsorbed flux in 1--10 keV band in the $RCS$ model.
The errors of the last five columns are estimated at the 90\% confidence level.
The observation 0550671101 suffered from severe contamination of flares
and is used only for determining $P$ and $\dot{P}$.
}
\enddata
  \tablenotetext{a}{\phantom{0}The exposure time of the 
flare-unscreened/flare-screened data.}
  \tablenotetext{b}{\phantom{0}The start epoch of the flare-unscreened data
after barycentric correction.}
  \tablenotetext{c}{\phantom{0}The 1$\sigma$ uncertainty 
\citep{1987A&A...180..275L}
of the last two digits is given in parentheses.}
\label{T:xmmobs}
\end{deluxetable*}

The field around this source was covered on several occasions 
spanning 2004--2007 and in 2008 and 2009.
We selected 12 archival \XMMN\ observations carried out during 
2008--2009 (PI J. Halpern) for our detailed analysis (see the observation 
information in Table~\ref{T:xmmobs}).
Here we only use the data collected with EPIC-MOS2, which happened to 
cover \src\ during both the 2008 and 2009 observations.
All the MOS observations were carried out in the full frame mode with a
time resolution of 2.6~s.
After removing the time intervals and observations with heavy proton 
flares, the total effective exposure time for the MOS2 observations 
amounts to 273~ks.
We used the flare-unscreened data to analyze the periodicities and 
the spin-down rate, while the flare-screened data are used for other 
analyses.
The observation taken on 2009 March 25 (ObsID 0550671101)
suffered from contamination by proton flares during most of the
observation time and hence was only used for determining the periodicity 
and the spin-down rate.
While the source was under the detection limit of \XMMN\ in 2004--2007 
and up to the observation taken in 2007 March 20--21, 
it  brightened during the 2008 September 19 observation (see 
Figure~\ref{f:transient}), and continued being detectable until 2009, 
according to the 16 archival  \XMMN\ observations which covered the 
source.  This indicates that this is a variable or transient source.

We reduced the \XMMN\ date using the Science Analysis System software
(SAS).\footnote {http://xmm.esac.esa.int/sas/}
We then used the XRONOS 
(version 5.22) and 
XSPEC 
(version 12.7.1) 
packages in HEASOFT\footnote{http://heasarc.gsfc.nasa.gov/lheasoft/} 
software (version 6.12) for timing and spectral analysis, respectively.

\section{Timing Analysis}
We first converted the photon arrival times to the solar system 
barycenter using the source coordinates R.A.\ =$\RAdot{18}{52}{46}6$, decldot.\ =
$\decl{00}{33}{20}9$, and then searched in the 12 observations for 
a periodicity in the power spectrum ($powspec$; 0.3--10 keV).
As shown in Figure~\ref{f:fig2}(a), we found a periodicity 
at $P~=~11.56~\s$ with high significance.
To refine the estimated period, we used an epoch-folding method ({\it 
efsearch}) and searched periodicities around 11.56~s with a step of
$10^{-5}$~s.
We determined the best-fit period\footnote{A similar period is obtained 
using the $Z^2_1$ test 
\citep{1983A&A...128..245B} .} 
and uncertainty in each observation (listed in Table~1) by using a least 
squares fit of a Gaussian to the 
observed $\chi^2$ value versus the period 
\citep{1987A&A...180..275L}.

We subsequently constructed a phase-connected timing solution with the 
observations spanning seven months.
To determine the time-of-arrival (TOA), we folded the time series using 
a fixed period $P=11.55871\s$ for each piece of data in the 0.3--10 keV energy 
band.  Using the single period is possible in this case since it is consistent 
with the period measurements determined for all observations.
A sinusoidal profile generally fitted the pulse profiles well and 
was thus used to estimate the TOAs.
We first iteratively fitted the TOAs of the 12 observations to a 
linear ephemeris with the TEMPO2 
software.\footnote{http://www.atnf.csiro.au/research/pulsar/tempo2/}
The model refines the period to $P=11.55871315 (5) \s$,
and gives an rms residual of $\sim0.9\%P$ ($\chi^2/{\rm degreesoffreedom
(dof)}=25.4/10$).
Adding a quadratic term gives a period $P=11.5587126 (4)~\s$ for MJD 
54728.75 and a spin-down rate $\dot{P}=5.7\pm4.8 \E{-14}~\s\ps$ 
(1$\sigma$ uncertainty), which slightly improves the fit ($\chi^2/ 
{\rm dof}= 18.5/9$ and rms residual $\sim 0.8\%P$, see Figure~\ref{f:fig2}(b)).
Hereafter we use $\dot{P} < 1.1\E{-13}~\s \ps$ due to the large error 
range of the best-fit $\dot{P}$ value.

The period and the spin-down rate indicate a dipolar surface magnetic 
field $B=3.2\E{19} (P\dot{P})^{1/2}~<~3.6\E{13}$~G.
The characteristic age and the spin-down luminosity are $\tau_c=P/(2
\dot{P})~>~1.7$~Myr and $\dot{E}_{\rm rot}~=~3.95\E{46}\dot{P} 
P^{-3}~<~2.8\E{30} \erg \ps$, respectively.

\begin{figure}
\centerline{ {\hfil\hfil
\psfig{figure=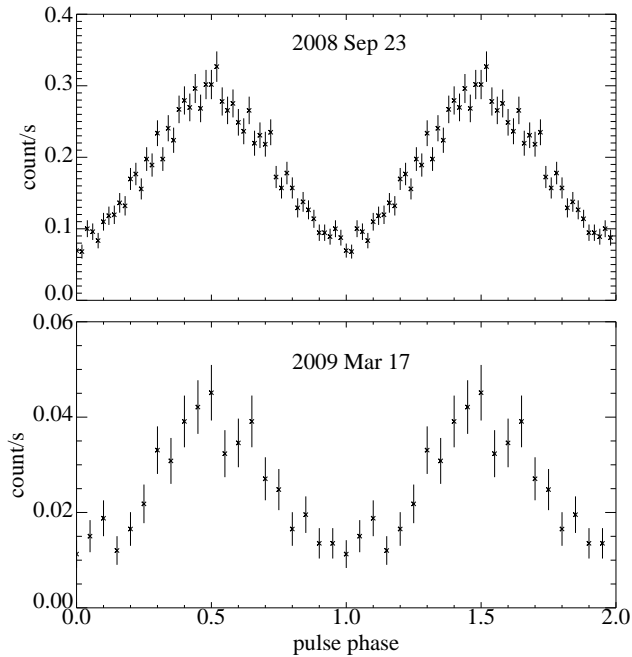,width=3.4in,angle=0, clip=}
\hfil\hfil}}
\caption{
Folded light curves of \src\ in the 0.3--10 keV band from two 
\XMMN\ observations.
}
\label{f:phase}
\end{figure}

\vspace{0.1in}

We folded the light curves in the 0.3--10 keV band with 50 bins/period and 
20 bins/period for the 2008 and 2009 observations (flare-screened),
respectively (see Figure~\ref{f:phase} for the pulse profiles).  
We also folded the light curves in the 0.3--2 keV and 2--10 keV bands to 
check for any variations, but we found no significant difference between the 
pulse profiles in the soft and hard bands.
The pulsed fractions $f_p$ after background subtraction are
$\sim56\%$--77\% (see Table~1).
Here $f_p$ is defined as the count rate ratio of the pulsed 
emission to the total emission, and the lowest bin in the folded light curve 
was taken to represent the unpulsed level.

\section{Spectral Analysis}
We extracted the 11 flare-screened EPIC-MOS2 spectra of \src\ from 
a circular region centered at the newly discovered source within a radius 
of $30''$ with the local background subtracted from a nearby source-free 
region. 
All spectra were then adaptively binned to reach a background-subtracted
signal-to-noise ratio of at least four.

We performed a joint-fit to the 11 spectra in the 0.5-10 keV energy range.
For the foreground interstellar absorption, we applied the $phabs$ model 
with the 
\citet{1989GeCoA..53..197A} abundances and the photo-electric 
cross-sections from 
\citet{1992ApJ...400..699B}.
We first adopt absorbed $power$-$law$ and $bbodyrad$ (a blackbody model) 
models to fit the spectra, respectively, with a common absorption 
column density $\NH$ for all spectra.
The $power$-$law$ model does not reproduce the spectra well ($\chi_\nu^2$
(dof)=1.46 (893)) with the photon index ranging from 3.1 to 3.9.
The single $bbodyrad$ model with the foreground absorption $\NH=1.36\pm0.05
\E{22}\cm^{-2}$ provides a good fit ($\chi_\nu^2$~(dof)= 1.10~(893)), 
noting that it under-estimates the hard X-ray 
tail above $\sim$6~keV.
Adding a power-law or a blackbody component was not however statistically needed,
given that only a few bins are above 6 keV and in the two-component model 
($blackbody+blackbody$ or $blackbody+power$-$law$) the parameters of the 
hard component were not well constrained.

\begin{figure}
\centerline{ 
\psfig{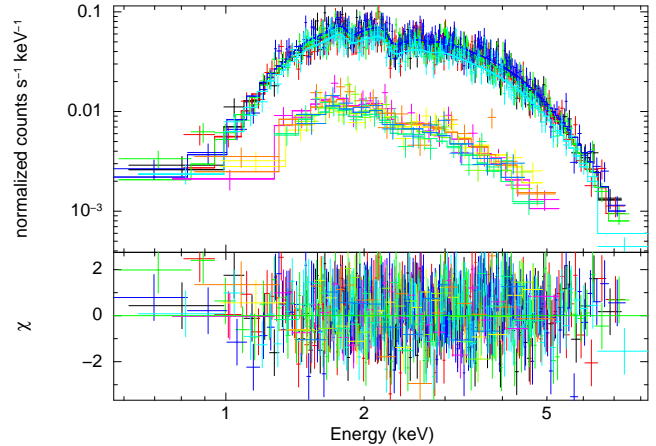} }
\caption{
\XMMN\ EPIC-MOS2 spectra of the 11 observations (see Table~1) fitted 
by the RCS model,
colored black, red, green, blue, light blue, magenta, yellow, orange,
yellow+green, green+cyan, blue+cyan, and blue+magenta in the sequence of 
the observation ID.
The upper and lower spectra are from the 2008 and 2009 observations,
respectively. See Table~1 for a summary of the observations shown
and the corresponding spectral parameters.
}
\label{f:spec}
\end{figure}

We subsequently applied the resonant cyclotron scattering (RCS) model 
\citep{2008ApJ...686.1245R, 2006MNRAS.368..690L},
which accounts for RCS of the thermal 
surface emission by the hot magnetospheric plasma.
This model has been successfully applied to a sample of magnetar spectra 
\citep{2008ApJ...686.1245R}.
The RCS model with $\NH=1.51\pm0.07\E{22} \cm^{-2}$ (constrained to be the
same in all the observations) provided the best fit to the spectra 
($\chi_\nu^2$~(dof)=1.07~(891), the optical depth in the scattering slab
$\tau_{\rm res}=1$--3 and thermal velocity of the magnetospheric electrons
$\beta=0.31^{+0.08}_{-0.15} c$), 
suggesting a magnetar nature of the source.  
Figure~\ref{f:spec} shows the spectra fitted with this model, and
Table~1 summarizes the best fit results of the $power$-$law$, $bbodyrad$ 
and RCS models, including the power-law photon index $\Gamma$, the blackbody 
temperature $kT_{bb}$ and the corresponding radiating radius $R_{bb}$ 
inferred from the $bbodyrad$ model,
the temperatures $kT_{rcs}$ and the 1--10 keV unabsorbed X-ray fluxes 
$F_X$ estimated from the best fit $RCS$ model.

As shown in Table~1, the X-ray properties of \src\ varied over the 
seven month timescale covered by the 11 observations.
The flux decreased by a factor of $\sim$6, along with a decrease of
the temperature from $\sim0.8$~keV to $\sim0.6$~keV according to the 
$bbodyrad$ model, and from $\sim 0.7$~keV to $\sim0.5$~keV based on 
the RCS model.
Using the blackbody model, we find that the size of the emitting area,
$R_{\rm bb}$, varied between $\sim0.7$~km and $\sim0.4$~km (at a distance of
7.1~kpc, see details in Section~5), a value much 
smaller than the typical radius of a neutron star ($\sim 10$~km).
This indicates that the blackbody X-ray emission is emitted from a small 
area (e.g., a hot spot) rather than from the whole surface of the neutron 
star. 

\section{Discussion}
We discovered a new transient 11.56-s X-ray pulsar coincident with the 
point source \src\ in the \XMMN\ Serendipitous Source Catalogue and
located south of SNR~\snr.
In this section, we further point out its low-B magnetar nature.

The previous \XMMN\ observations in 2004 and 2007 did not detect \src.
Neither was it detected in an archival \Chandra\ observation taken in 
2001 (ObsID 1982), nor in the \Chandra\ source catalog (release 
1.1).\footnote{http://cxc.harvard.edu/csc/}
Also, it was not detected by \ROSAT\ observations taken in 
1996--1997, and does not appear in the WGA Catalog of \ROSAT\ point 
sources.\footnote{http://heasarc.gsfc.nasa.gov/wgacat/}
We also checked all the archival \Swift\ observations of 3XMM~J185246.9+003318 
taken from 2012 September 25 to 2013 September 27, and we do not detect
the source with either the X-Ray Telescope or the Burst Alert Telescope.
Furthermore, we have not found any optical or infrared counterpart within 
$2\farcs4$ (the 1$\sigma$ positional uncertainties for 90\%
of the points sources in the 3XMM catalog) radius of the source in the Digital 
Sky Survey-2, the AllWISE Source Catalog, the Two Micron All Sky Survey images 
and All-Sky Point Source Catalog, and the GLIMPSE I Spring '07 Catalog and 
Archive.

The foreground absorption given by the best-fit RCS model, $\NH\simeq 
1.5\E{22} \cm^{-2}$, is consistent with that of SNR \snr\ 
(1.54--1.78 $\E{22} \cm^{-2}$, \citealp{2004ApJ...605..742S};
1.50--1.53 $\E{22}\cm^{-2}$, \citealp{2009A&A...507..841G}),
suggesting that \src\ is likely located at a distance similar to that of \snr.
Hence, we adopt a distance $d$ of 7.1~kpc toward \src,
the same as that inferred for Kes~79 
\citep{1998ApJ...504..761C, 1989ApJ...336..854F},
and subsequently parameterize the physical properties 
with $d$/7.1~\kpc.

The data show a spectral evolution of the new source.
The average X-ray flux of \src\ in the 1--10 keV band was around $4.5
\E{-12} \erg\ps$ in 2008 and decreased to around $7.7\E{-13} \erg\ps$ in 2009.
The corresponding luminosities are thus inferred to be $L_X\approx2.7\E{34} 
(d/7.1~\kpc)^2$ $\erg\ps$ and $4.6\E{33}(d/7.1~\kpc)^2$~$\erg\ps$, 
respectively, both of which rule out rotational energy loss
($<2.8\E{30}~\erg\ps$) as the source of the X-ray emission.
This is one of the defining properties of magnetars.
Along with the variation of the luminosity, the temperature
$kT_{\rm bb}$ and the X-ray emitting radius $R_{\rm bb}$ in the blackbody
scenario decreased as the flux decreased.
The softening of the X-ray emission has been observed in transient magnetars 
during the outbursts decay 
\cite[see, e.g.,][]{2011arXiv:1101.4472}.

We have thus provided firm evidence for \src\ being a new, transient magnetar,
namely the slow pulsations, the X-ray luminosity higher than the spin-down 
power, the X-ray spectra characterized by a blackbody/RCS model that softened 
as the luminosity decreased, the lack of detection of any optical or infrared 
counterpart, and non-detection in archival 
X-ray observations prior to the 2008 September 19 \XMMN\ observation.

With a period of 11.5587126 (4)~s (at MJD 54728.75), \src\ has the second 
longest period among currently known magnetars 
\citep[after AXP~1E~1841$-$045 with $P=11.78$~s;][]{1997ApJ...486L.129V}
and the longest period among the nine known transient magnetars.
The low dipolar magnetic field $B<3.6\E{13}$~G deduced from the period
and period derivative suggests that this is the third low-B magnetar 
discovered so far, joining SGR~0418$+$5729 and Swift~J1822.3$-$1606.

Given that the source was not detected during the 2007 March 20 \XMMN\
observation but clearly detected in the 2008 September 19 observation,
we conclude that an outburst likely occurred between these two periods.
It is unclear when the outburst of \src\ started prior to the 2008 
observation and whether the source was entering the quiescent stage 
during 2009.
Since the spectrum of \src\ during the 2008--2009 observations did not require 
an additional power-law or hotter blackbody component, as is commonly 
observed (although with a few exceptions) in magnetars in quiescence,
including 1E~1841$-$045 
\citep{2010ApJ...725L.191K, 2013arXiv1309.4167O}
and since the source is currently not detected by \textit{SWIFT},
the source should remain in the outburst decay phase during the \XMMN\ 
observations reported here. 
Indeed, the blackbody component dominates the emssion for a handful of 
magnetars during the later phases of the outburst decay 
\citep{2011arXiv:1101.4472}.
Moreover, the non-detection of \src\ in the X-ray observations before 2007 
March 20 supports that the source was in quiescent state.
The non-detection of it by \Chandra\ ACIS-I during 2001 sets the 
$3\sigma$ upper limit of the source's count rate to $1\E{-3}$ counts$\ps$
(1--10 keV) , which corresponds to an unabsorbed flux $6\E{-14}\erg 
\cm^{-2}\ps$ (assuming $\NH=1.5\E{22}\cm^{-2}$ and a temperature $kT_{\rm bb}=
0.1$~keV that is similar to that inferred for the low-B magnetar 
Swift~J1822.3$-$1606 during quiescence; \citealp{2012ApJ...761...66S}; 
a lower flux would be obtained if the quiescent temperature is larger).
The luminosity of \src\ in its quiescent stage is thus roughly estimated to 
be $<4\E{32}~\erg\ps$ (1--10~keV), which is a factor of over 68
lower than that during 2008.
This significant flux variation was commonly seen in several other 
magnetar outbursts.
One explanation of the outburst involves the fast release of the energy 
in the crust 
\citep{2002ApJ...580L..69L}.
Another explanation is given by the untwisting magnetosphere model,
which predicts a hot spot forms at the foot-prints of the current-carrying 
bundle of field lines, and the spot shrinks with the decrease of luminosity 
and temperature 
\citep{2009ApJ...703.1044B}.
It can interpret the decreased emitting area 
($R_{\rm bb}$ from $\sim 0.7$~km to $\sim0.4$~km) along with the spectral 
evolution found in this study.

It is well known that a 105 ms X-ray pulsar CXOU 185238.6$+$004020
\citep{2003ApJ...584..414S}
is located at the center of SNR~\snr, and is identified as
an ``anti-magnetar" with a dipole magnetic field $B = 3.1\times 10^{10}$~G
\citep{2005ApJ...627..390G, 2007ApJ...665.1304H, 2010ApJ...709..436H}.
Considering the similar $\NH$ inferred for both \src\ and \snr,
it would be interesting to explore whether CXOU 185238.6$+$004020 or \src\ 
is associated with Kes 79.
Note that in both cases the two objects could once be in a binary
system or have no relation with each other. 
Given a dynamical age $t_{\rm SNR}\sim5.4$--7.5 kyr 
\citep{2004ApJ...605..742S}
or a Sedov age $\sim 5$ kyr (P. Zhou \etal, in preparation) for SNR~\snr,
if \src\ was formed from the supernova that 
produced Kes 79, its projected velocity would be $\sim 3\E{3}
(t_{\rm SNR}/5$~kyr)$^{-1}$~\kms.  
This velocity is very high when compared to the mean measured velocity
of a sample of six magnetars ($200\pm 90$~\kms; 
\citealp{2013ApJ...772...31T}).
Nevertheless, a high velocity has been proposed for another
magnetar (1100~\kms\ for SGR 0526$-$66 if associated with SNR N49;
\citealp{2012ApJ...748..117P}), 
and there is accumulating evidence for high pulsar speed,
even exceeding $4\E{3}$~\kms\ 
\citep[e.g.,][]{2005MNRAS.362.1189Z}.
However, the magnetar has a characteristic age $\tau_c >1.7$~Myr,
and the low quiescent luminosity ($<4\E{32}\erg\ps$, assuming $kT_{\rm bb}$=0.1 
keV) suggests a large age of the magnetar (0.1--1~Myr) estimated from a 
magneto-thermal evolution model 
\citep{2013MNRAS.434..123V},
which makes the association rather unlikely.
Future monitoring observations and a proper motion measurement
of \src\ are needed to confirm or refute this interesting scenario.

{\it Note added in manuscript.}---Fifteen days after our Letter was posted 
on arXiv  on 2013 October 29 and while it was under revision, 
\citet{2013arXiv1311.3091R}
posted a Letter on this source.
They independently confirmed our findings and the low-B magnetar nature of 
this source.

\begin{acknowledgements}
We gratefully acknowledge the anonymous referee for suggestions regarding 
phase-connected timing analysis and for comments that helped improve
the Letter.
We thank Tomaso Belloni, Michael Nowark, Wen-Fei Yu, Fang-Jun Lu, and
Zhong-Xiang Wang for helpful discussions.
We also thank Nanda Rea, Shuang-Nan Zhang, Ren-Xin Xu, Jin-Lin Han, Na Wang, 
Thomas Tauris, and Shriharsh P. Tendulkar for valuable comments,
Song Huang for checking any optical/infrared counterpart,
and Harsha Kumar for checking the \Swift\ observations.
P.Z. is grateful to the COSPAR Capacity Building Workshop, 2013.
Y.C. and X.D.L. are grateful for the support from 
the 973 Program grant 2009CB824800, 
NSFC grants 11233001, 11133001, and 11333004,
the grant 20120091110048 by the Educational Ministry of China, 
and the Qinglan Project of Jiangsu Province.
S.S.H. acknowledges support from  NSERC through the Canada Research Chairs
Program and a Discovery Grant, and from the Canadian Space Agency, 
the Canadian Institute for Theoretical Astrophysics, and Canada's Foundation 
for Innovation.
\end{acknowledgements}

\end{document}